\documentclass[conference]{IEEEtran}
\IEEEoverridecommandlockouts
\usepackage{cite}
\usepackage{amsmath,amssymb,amsfonts}
\usepackage{algorithmic}
\usepackage{graphicx}
\usepackage{tabularx}
\usepackage{textcomp}
\usepackage{xcolor}
\usepackage{stfloats}
\usepackage{multirow}
\def\BibTeX{{\rm B\kern-.05em{\sc i\kern-.025em b}\kern-.08em
    T\kern-.1667em\lower.7ex\hbox{E}\kern-.125emX}}
\begin{document}
	\title{A 3D Non-Stationary Channel Model for 6G Wireless Systems Employing Intelligent Reflecting Surface}
	\author{Yingzhuo Sun\textsuperscript{1,2}, Cheng-Xiang Wang\textsuperscript{1,2,*}, Jie Huang\textsuperscript{1,2}, and Jun Wang\textsuperscript{1,2}
		\\
		\textsuperscript{1}National Mobile Communications Research Laboratory, School of Information of Science and Engineering, 
		\\Southeast University, Nanjing 210096, China.\\
		\textsuperscript{2}Purple Mountain Laboratories, Nanjing 211111, China.\\
		\textsuperscript{*}Corresponding Author: Cheng-Xiang Wang
		\\
		Email: \{sunyingzhuo, chxwang, j\_huang, jun.wang\}@seu.edu.cn
	}
	\maketitle
\begin{abstract}
	As one of the key technologies for the sixth generation (6G) mobile communications, intelligent reflecting surface (IRS) has the advantages of low power consumption, low cost, and simple design methods. But channel modeling is still an open issue in this field currently. In this paper, we propose a three-dimensional (3D) geometry based stochastic model (GBSM) for a massive multiple-input multiple-output (MIMO) communication system employing IRS. The model supports the movements of the transmitter, the receiver, and clusters. The evolution of clusters on the linear array and planar array is also considered in the proposed model. In addition, the generation of reflecting coefficient is incorporated into the model and the path loss of the sub-channel assisted by IRS is also proposed. The steering vector is set up at the base station for the cooperation with IRS. Through studying statistical properties such as the temporal autocorrelation function and space correlation function, the non-stationary properties are verified. The good agreement between the simulation results and the analytical results illustrates the correctness of the proposed channel model.
\end{abstract}
\section{Introduction}
	With the advent of a new information era, larger volume of data, higher transmission rate, and safer communication are appealed by users \cite{You1},\cite{6Gchannel}. To fulfill these demands, there are many new solutions and technologies arising recently. IRS draws much attention due to its unique advantages. At the beginning of the development of wireless communication, the environment is always considered as the object to compete with. Researchers are devoted to designing the transceiver structure and the complex algorithms of signal processing in order to decrease the negative effect of the environment. But the core idea of IRS is to change and control the communication environment \cite{You1}. It is envisioned as a key technology of 6G because of its low power consumption and low cost. 
	
	Recent research directions on IRS mainly consist of the optimization design of the reflecting coefficient, channel estimation, application in physical security, etc. Reference \cite{Huang1} considered the downlink transmission scenario assisted by IRS and optimized the precoding matrix and the phase shift matrix simultaneously. The maximization of system energy efficiency under the condition of finite resolution was also investigated in \cite{Huang2}. Authors in \cite{Huang3} studied the simulation results about the energy efficiency and the system throughput according to more realistic system settings and parameters. In \cite{WQ1}\cite{WQ2}, authors jointly designed the beamforming at a base station and IRS in order to minimize the transmitting power. In \cite{WQ3}, they further studied the same optimization problem considering the discrete phase shift. The influence on system performance caused by the phase error was studied in \cite{PERROR}. The authors in \cite{SE} gave a thought of a joint optimization problem on spectrum efficiency. Different conditions of Rician factors and discrete phase shift were studied in \cite{Jinshi}. 
	
	On channel estimation, researchers in \cite{Chesti1} proposed a new channel estimation protocol for the multiple-input single-output (MISO) system employing IRS. To reduce the training overhead, the authors in \cite{Chesti2} considered two independent methods, one is based on compressed sensing, the other is based on deep learning. In this work, a new IRS structure based on channel sensors was proposed. In \cite{Chesti3}, the authors studied the problem of estimating the IRS-assisted cascaded channel with multi antennas. 
	
	Physical layer safety on IRS was investigated in \cite{Physafe1}. For the application of using IRS as transceivers, researchers focused on output probability \cite{transrece1}, asympototic data rate \cite{transrece2}, and uplink spectrum efficiency \cite{transrece3}. Because of the importance of high frequency communication in the fifth generation (5G) mobile communications \cite{5GCMsur}, there are also researchers interested in high frequecy band technology combined with IRS such as millimeter wave (mmWave) and terahertz (THz) communication \cite{highband1}, visible light communication \cite{highband2}. Orthogonal frequency division multiple access (OFDM) incorporated with IRS was studied in \cite{OFDM}. Reference \cite{MIRS} analyzed the system with multiple IRS. 
	
	To the best of our knowledge, there is little research achievement on channel modeling with IRS. In this paper, a novel channel model with IRS is proposed. The overall idea of this article is to split the total channel into three sub-channels and consider large scale fading and small scale fading, respectively. At the same time, the steering vector for the base station and the reflecting coefficient matrix for IRS are generated. 
	
	The remainder of this paper is organized as follows. Section~\uppercase\expandafter{\romannumeral2} describes the channel model in detail. In Section~\uppercase\expandafter{\romannumeral3}, the statistical properties of the proposed channel model are calculated. The simulation results are presented in this section. Finally, we draw conclusions in Section~\uppercase\expandafter{\romannumeral4}.
	
\section{A Novel 3D IRS MIMO GBSM}
\subsection{Description of the Channel Model}
\subsubsection{The Framework of the Channel Model}
	There are many right occasions for IRS, such as low received power caused by blockages, physical layer safety and so on. In this article, we choose the commonest occasion which is using IRS to overcome the dead zone effect. The scenario is illustrated in Fig. 1. BS-IRS channel, IRS-USER channel, and BS-USER are three sub-channels of the whole system. The channel coefficientS matrices of them are denoted as $\textbf{H}_{BI}$, $\textbf{H}_{IU}$, and $\textbf{H}_{BU}$, respectively. The capital form of $\textbf{h}$ means that the channel coefficient in the matrices consists of large scale fading. The channel coefficient matrices that only consist of small scale fading are denoted as $\textbf{h}_{BI}$, $\textbf{h}_{IU}$, and $\textbf{h}_{BU}$, respectively. The reflecting coefficient matrix of IRS is presented as $\mathbf{\Phi}$. At last, the overall channel coefficient matrix is illustrated as $\textbf{H}_\text{total}$. 
	
\begin{figure}[tb]
	\centerline{\includegraphics[width=0.5\textwidth]{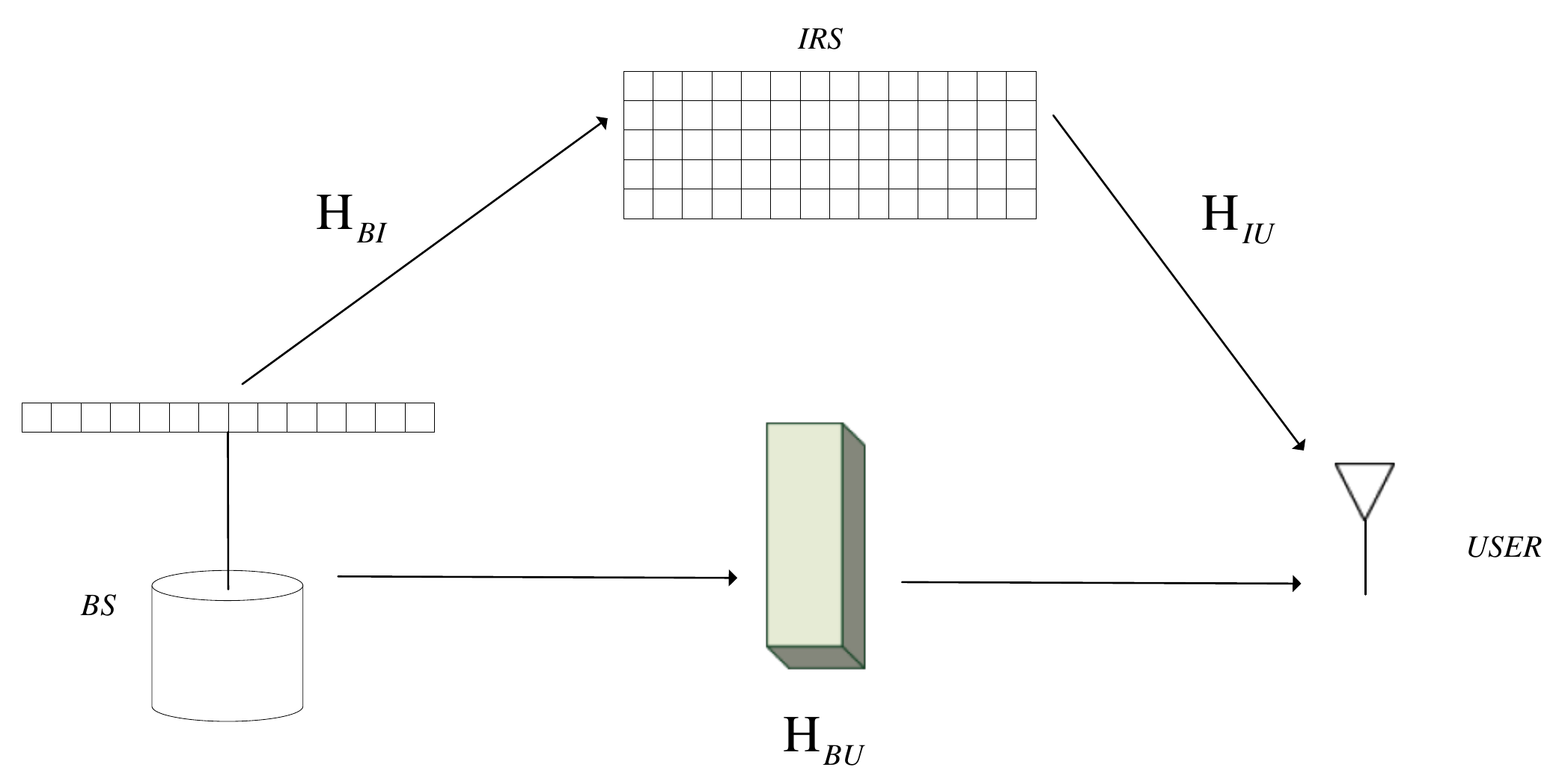}}
	\caption{A wireless communication scenario using IRS.}
	\label{fig_1}
\end{figure}	
	
	 We consider two kinds of large scale fading, the first one is shadowing fading, another is path loss. We use $SF_{BI}$, $SF_{BU}$, and $SF_{IU}$ to present the corresponding lognormal random variables of different subchannels to model the shadowing fading effect. $PL_{BU}$ is the path loss of the subchannel which is between BS and user. $PL_{BIU}$ is the path loss of the cascaded channel assisted by IRS. $\textbf{f}$ is the steering vector of BS. Then, the channel coefficient matrix can be expressed as
\begin{equation}
\begin{split}
	\textbf{H}_\text{total}&=(\textbf{H}_{IU}\mathbf{\Phi}\textbf{H}_{BI}+\textbf{H}_{BU})\textbf{f}\\
	&=(\sqrt{SF_{BI}SF_{IU}PL_{BIU}}\textbf{h}_{BI}\mathbf{\Phi}\textbf{h}_{IU}\\
	&+\sqrt{SF_{BU}PL_{BU}}\textbf{h}_{BU})\textbf{f}.
\end{split}
\end{equation}

\subsubsection{The Reflecting Coefficient Matrix Setting}
	In this part, we will discuss how to achieve the reflecting coefficient matrix of IRS. Firstly, we consider a scenario as shown in Fig. 2. The transmitter ($\text{T}_\text{X}$) and the receiver ($\text{R}_\text{X}$) are both employed with one antenna. Assume that the directions of peak radiation of the transmitting and receiving antennas point to the center of IRS. The ordered pair $(x,y)$ means the index of each element on IRS. The distances between $\text{T}_\text{X}$ and each element on IRS are denoted as $r_{x,y}^t$. The distances between $\text{R}_\text{X}$ and each element are presented as $r_{x,y}^r$. $d_x$ and $d_y$ are intervals of IRS on two directions. $\phi_{x,y}$ is the phase of corresponding element on IRS. $M_x$ and $M_y$ are the number of elements on horizontal and vertical directions of IRS. Here we assume that $M_x$ and $M_y$ are both even. Given these premises, we can get the relationship between the received power and the transmitted power as\cite{TangWK}
\begin{equation}\label{ptpr}
\begin{split}
P_r=&P_t\frac{d_{x}d_{y}\lambda^2}{64{\pi}^3}\\
&\left|\sum_{x=1-\frac{M_x}{2}}^{\frac{M_x}{2}}\sum_{x=1-\frac{M_y}{2}}^{\frac{M_y}{2}}\frac{e^{\frac{-j(2\pi(r_{x,y}^r+r_{x,y}^t)-\lambda \phi_{x,y})}{\lambda}}}{r_{x,y}^{r}r_{x,y}^t}\right|^2.
\end{split}
\end{equation}
\begin{figure}[tb]
	\centerline{\includegraphics[width=0.5\textwidth]{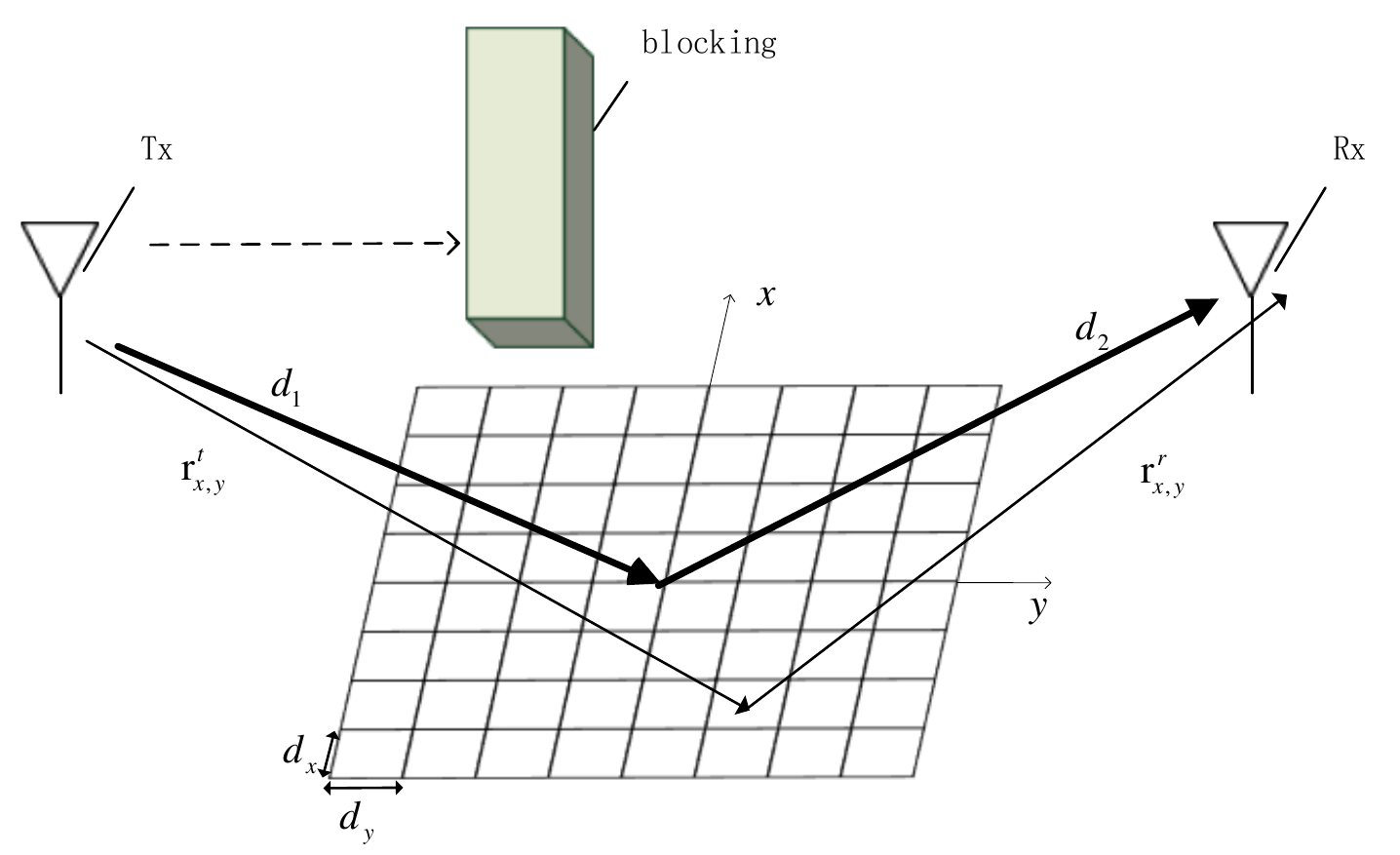}}
	\caption{A scenario for deriving reflecting coefficients.}
	\label{fig_3}
\end{figure}

	Then, we can easily see that the received power is maximized when the phase shifts fulfill $\phi_{x,y}=\text{mod}(\frac{2\pi(r_{x,y}^t+r_{x,y}^t)}{\lambda},2\pi)$. Here, mod means the modulo operation. So the reflecting coefficients matrix can be obtained according to the line. Here we use one index $r$ to replace $(x,y)$ to find one element. The relationship between them is $r=(x-1)M_{y}+y$. The matrix is thus shown as
\begin{equation}
	\mathbf{\Phi}=\text{diag}(e^{j\phi_1},e^{j\phi_2}...e^{j\phi_r}...e^{j\phi_{M_x\times M_y}}).
\end{equation}
	where $\text{diag}(\cdot)$ means the operation of generating a diagonal matrix using the numbers in the embrace as its elements.
	
	In the following part of this article, we will consider the situation of multiple antennas. To continue using the former conclusion of reflecting coefficients, the only thing we should do is to ensure that the peak radiation of multiple antennas is still pointing to the center of IRS. We will use a steering vector for the transmitter. Assume that all the antennas are omnidirectional. According to \cite{steering}, we can obtain the steering vector expressed as
\begin{equation}\label{steering}
	\mathbf{f}=\left[c_{1}(\mathbf{\Omega}),c_{2}(\mathbf{\Omega})...c_{m}(\mathbf{\Omega})...c_{M_T}(\mathbf{\Omega})\right].
\end{equation}

	In (\ref{steering}), $\mathbf{\Omega}$ means the direction pointing to the center of IRS, and $c_{m}(\mathbf{\Omega})$ is the corresponding coefficient of the $m\text{th}$ antenna. It is calculated as \cite{steering}
\begin{equation}\label{steering2}
	c_{m}(\mathbf{\Omega})=\text(exp)(j2\pi \lambda^{-1} \left\langle\textbf{e}(\mathbf{\Omega}),\textbf{r}_m\right\rangle  +j2\pi\nu_{l}t)
\end{equation}
	where $\nu_{l}$ is the doppler shift. $\textbf{e}(\mathbf{\Omega})$ and $\textbf{r}_m$ are the unit vector of the departure direction and the vector related to the antenna interval which is calculated as $\left(x_{m}-d_{x},y_{m}-d_{y},z_{m}-d_{z}\right)$. $(x_{m},y_{m},z_{m})$ is the Cartesian value. $d_{x}$, $d_{y}$ and $d_{z}$ are usually valued as a half of the wavelength. Through substituting the optimized reflecting coefficients into (\ref{ptpr}), we can get the path loss of the subchannel assisted by IRS.
	
\subsubsection{Small Scale Fading Channel Coefficient Matrix}
	Geometry based stochastic model (GBSM) is adopted here to generate small scale fading channel coefficients. Firstly, consider a MIMO system. We assume that all the antennas are omnidirectional. For each of three subchannels, the methods of generating the small scale fading coefficient are similar. So we choose one subchannel to illustrate the method. The GBSM is shown in Fig. 3. The interval distances of IRS on the $x$ and $y$ direction are denoted as $\delta^{I_x}$ and $\delta^{I_y}$, respectively. The interval of IRS is denoted as $\delta^{B}$. $\beta_A^B$ is the azimuth angle of BS array. $\beta_E^B$ is the elevation angle of BS array. $\beta_A^{I_x}$ and $\beta_A^{I_y}$ are azimuth angles of two extending directions of IRS. Two elevation angles of two extending directions of IRS are denoted as $\beta_E^{I_x}$ and $\beta_E^{I_y}$. The $q\text{th}$,$r\text{th}$ and $p\text{th}$ element of BS, IRS and user are presented as $A_q^B$,$A_r^I$ and $A_p^U$. The total number of paths between two corresponding elements of BS and IRS is denoted as $N_{qr}(t)$. To simplify the illustration, the figure only shows one path. The first bounce cluster and the last bounce cluster of the $n\text{th}$ path for BS-IRS subchannel are denoted as $C_n^{A,BI}$ and $C_n^{Z,BI}$. The propagation between them is illustrated as a virtual link.
	
	The speed vectors of BS, user and clusters at different time instance are denoted as $\mathbf{v}^{B}(t)$, $\mathbf{v}^{U}(t)$, $\mathbf{v}^{A_n}(t)$ and $\mathbf{v}^{Z_n}(t)$. We assume that they move in a 2D plane. So just four angles can determine the directions of their movements which are presented as $\alpha^{B}(t)$, $\alpha^{U}(t)$, $\alpha^{A_n}(t)$, $\alpha^{Z_n}(t)$.
	
	Let $\phi_{A,m_n}^{B,BI}$ and $\phi_{E,m_n}^{B,BI}$ be azimuth angle of departure (AAoD) and elevation angle of departure (EAoD) of the $m\text{th}$ ray at initial time. Similarly, let $\phi_{A,m_n}^{I,BI}$ and $\phi_{E,m_n}^{I,BI}$ be the azimuth angle and the elevation angle of the $m\text{th}$ ray coming from $C_n^{A,BI}$ received by $A_1^I$. The scatterer interacting with the $m\text{th}$ ray in the $n\text{th}$ cluster is denoted as $S_{m_n}^{A,BI}$ or $S_{m_n}^{Z,BI}$. The distances between each antenna element and each scatterer are denoted as $d_{q,m_n}^{B,BI}(t)$,$d_{r,m_n}^{I,BI}(t)$. The distances between the first element and the corresponding scatterer at initial time are presented as $d_{m_n}^{B,BI}$, $d_{m_n}^{I,BI}$. 
	
	Then we will illustrate how to generate the small scale fading channel coefficients. The dimension of $\textbf{H}_{BI}$ is $M_{xy}\times M_{B}$. The elements of this matrix are denoted as $h_{qr}(t,\tau)$ that means the channel coefficient between $A_q^B$ and $A_r^I$ at time instant $t$. It is calculated as
\begin{equation}\label{channelresponse}
	h_{qr}(t,\tau)=\sqrt{\frac{K_R}{K_{R}+1}}h_{qr}^{L}(t,\tau)+\sqrt{\frac{1}{K_{R}+1}}h_{qr}^{N}(t,\tau).
\end{equation}

	It is obtained by summing two weighed components. The former one is the line of sight (LoS) component, and the other one is non-line of sight (NLoS) component. Their weights are determined by the Rician factor $K_R$. First, we calculate the NLoS component as\cite{Bianji}
\begin{equation}
\begin{split}\label{NCIR}
 h_{qr}^{N}(t,\tau)=&\sum_{n=1}^{N_{qr}(t)}\sum_{m_{n}=1}^{M_n}\sqrt{P_{qr,m_n}(t)}\\
  &e^{j2\pi f_{c}\tau_{qr,m_n}(t)}\cdot\delta(\tau-\tau_{qr,m_n}(t)).
\end{split}
\end{equation}

	Here, we consider the condition of omnidirectional antennas. So, the antenna patterns are ignored. In (\ref{NCIR}), $P_{qr,m_n}(t)$ and $\tau_{qr,m_n}(t)$ are the power and delay of the $n\text{th}$ pair of clusters between $A_q^B$ and $A_R^I$ at time instant $t$, respectively. The delay is calculated as\cite{Bianji}
\begin{equation}\label{nlosdelay}
	\tau_{qr,m_n}(t)=d_{qr,m_n}(t)/c+\tau_{m_n}^{v}.
\end{equation}

	In (\ref{nlosdelay}), $c$ means the speed of light. $\tau_{m_n}^{v}$ is the delay between two scatterers, $S_{m_n}^{A,BI}$ and $S_{m_n}^{v}$, which is modeled as $\tau_{m_n}^{v}=\tau_{\text{link}}+d_{m_n}/c$. $d_{m_n}$ is the distance between two scatterers. $\tau_{\text{link}}$ is a random variable that follows an exponential distribution. $d_{qr,m_n}(t)$ is calculated as $d_{qr,m_n}(t)=\left|\left|\textbf{d}_{q,m_n}^{B,BI}(t)\right| \right|+\left|\left| \textbf{d}_{r,m_n}^{I,BI}(t) \right| \right|$, where $||\cdot||$ means Frobenius norm. Two vectors are the vectors that corresponding antennas point to the scatterers. The former vector is calculated as\cite{Bianji}
\begin{equation}
	\textbf{d}_{q,m_n}^{B,BI}(t)=\textbf{d}_{m_n}^{B,BI}-[\textbf{l}_{q}^{B}+\int_{0}^{t}(\textbf{v}^{B}(t)-\textbf{v}^{A_n}(t))\,dt]
\end{equation}
	where $\textbf{d}_{m_n}^{B,BI}$ and $\textbf{l}_{q}^{B}$ are calculated as
\begin{equation}
	\textbf{l}_{q}^{B}=(q-1)\delta^{B}[\cos{\beta_E^B}\cos{\beta_A^B},\cos{\beta_E^B}\sin{\beta_A^B},\sin{\beta_E^B}].
\end{equation}

	In (\ref{channelresponse}), the LoS component is calculated as
\begin{equation}
	h_{qr}^{L}(t,\tau)=e^{j2\pi f_{c}\tau_{qr}^{L}(t)}\cdot\delta(\tau-\tau_{qr}^{L}(t))
\end{equation}
	where the $\tau_{qr}^{L}(t)$ is the propogation delay of LoS component varied with space and time. It is calculated as $\tau_{qr}^{L}(t)=D_{qr}(t)/c$. Here $D_{qr}(t)=||\textbf{D}_{qr}(t)||$ is the distance between $A_q^B$ and $A_r^I$. Then, the vector is calculated as \cite{Bianji}
\begin{equation}
	\textbf{D}_{qr}(t)=\textbf{D}_{BI}+\textbf{l}_q^{B}-\textbf{l}_r^{l}+\int_{0}^{t}\mathbf{v}^{B}(t)\,dt.
\end{equation}
\begin{figure}[tb]
	\centerline{\includegraphics[width=0.5\textwidth]{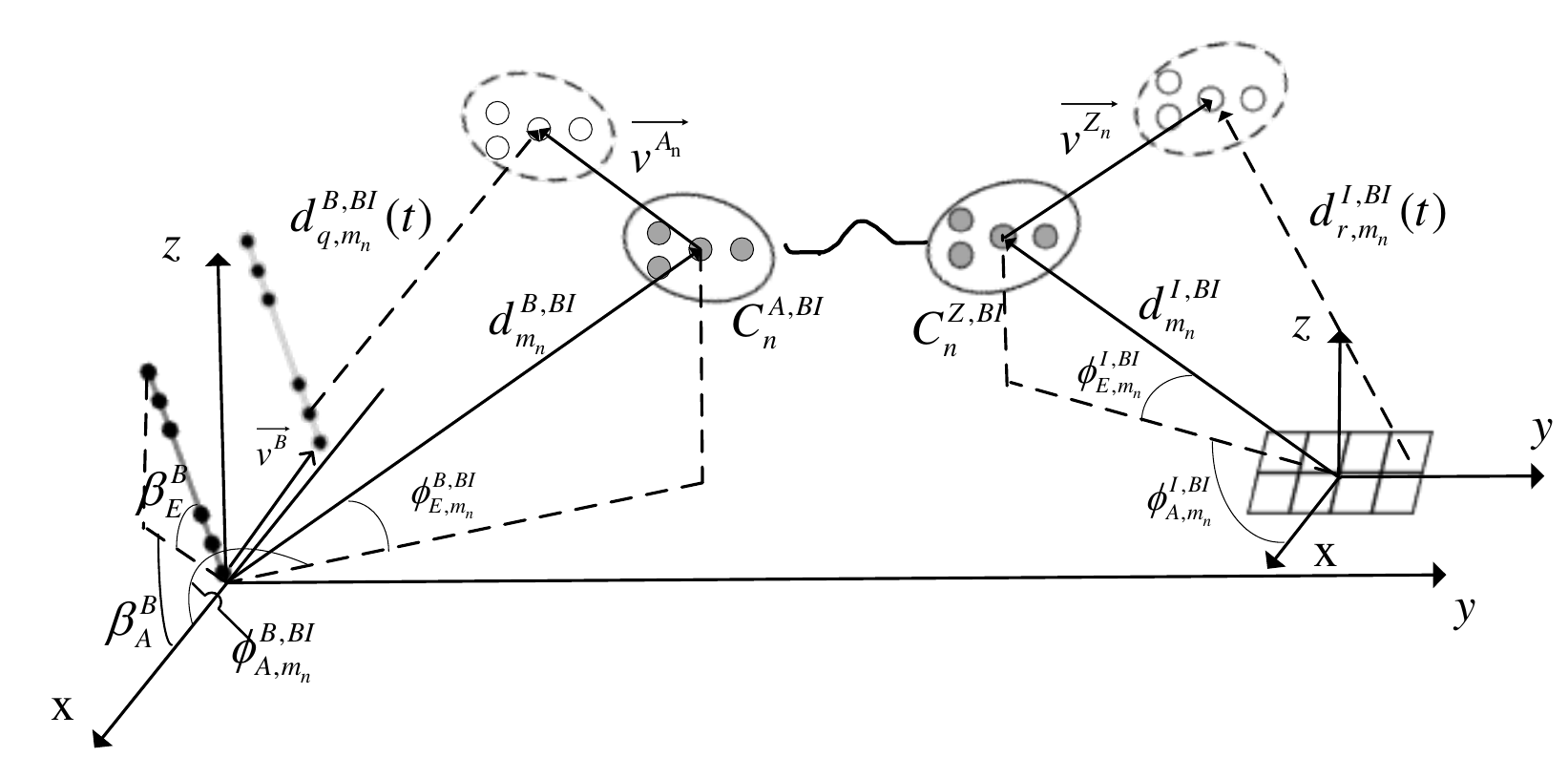}}
	\caption{The GBSM for IRS channel.}
	\label{fig_4}
\end{figure}
\subsubsection{The Distribution of Scatters in a Cluster}
	We define the coordinate that employs antenna array be a benchmark as Global Coordinate System(GCS) and the one that lets the center of a cluster be a benchmark as Local Coordinate System(LCS). The coordinate value of GCS is $(x,y,z,\phi,\theta)$. The value of LCS is $(x',y',z',\phi',\theta')$. The transformation between them is determined by three angles which are bearing angle $\alpha$, downtilt angle $\beta$ and slant angle $\gamma$. The transformation is based on the theory that any 3D rotation can be divided into three rotations around rotating axis when the order of rotating axises are determined. Under this condition, $\alpha$, $\beta$, and $\gamma$ are fixed. The relationship btween two sets of coordinate values is shown as \cite{3GPP}
\begin{equation}
	[x',y',z']=[x,y,z]\cdot\textbf{R}
\end{equation}
	where $R$ is the rotation matrix, which is calculated as \cite{3GPP}
\begin{equation}
\begin{split}
	\textbf{R}=\left[
	\begin{matrix}
	\cos{\alpha}&-\sin{\alpha}&0\\
	\sin{\alpha}&\cos{\alpha}&0\\
	0&0&1
	\end{matrix}
	\right]
	&\left[
	\begin{matrix}
	\cos{\beta}&0&\sin{\beta}\\
	0&1&0\\
	-\sin{\beta}&0&\cos{\beta}
	\end{matrix}
	\right]\\
	\left[
	\begin{matrix}
	1&0&0\\
	0&\cos{\gamma}&-\sin{\gamma}\\
	0&\sin{\gamma}&\cos{\gamma}
	\end{matrix}
	\right].
\end{split}
\end{equation}

	The method of generating the coordinate value of LCS is under the condition of the coordinate value fulfilling ellipsoid gaussian distribution which means the joint probability distribution function of $x',y',$ and $z'$ is \cite{Bianji}
\begin{equation}
	p(x',y',z')=\frac{exp\left(-\frac{x'^2}{2\sigma_{x}^2}-\frac{y'^2}{2\sigma_{y}^2}-\frac{z'^2}{2\sigma_{z}^2}\right) }{(2\pi)^{\frac{3}{2}}\sigma_x\sigma_y\sigma_z}.
\end{equation}
\begin{figure}[tb]
	\centerline{\includegraphics[width=0.5\textwidth]{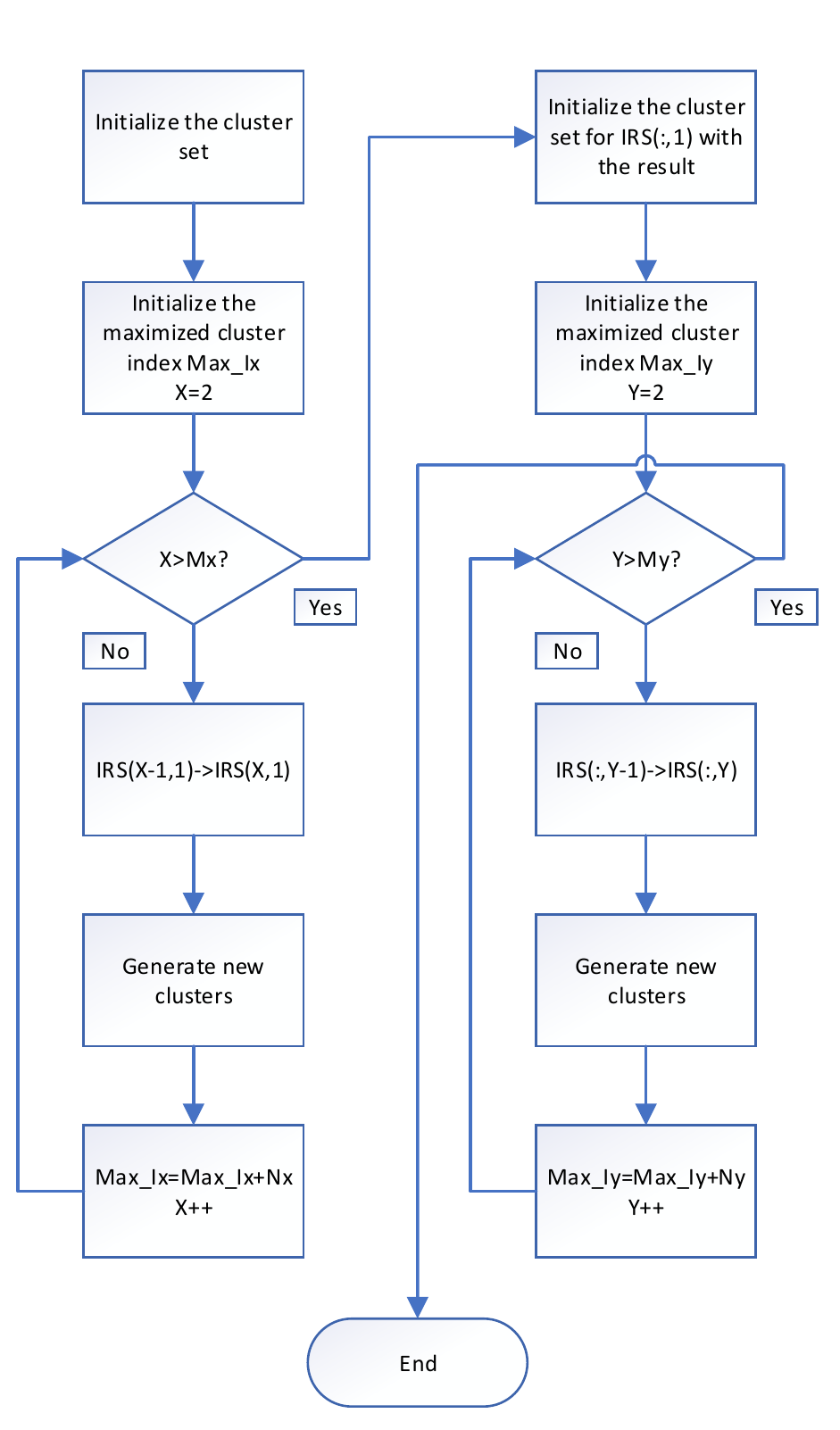}}
	\caption{The evolution procedure of clusters on the 2D plane.}
	\label{fig_5}
\end{figure}
\subsubsection{Cluster Evolution in Space Domain}
	Under the condition of using the large antenna array, it will bring the space non-stationarity and the clusters will evolve on the array. The basic idea is finishing one direction first and using the first evolution result as the initial state of the second one. Here the evolution on one direction is mainly discussed. The death rate and the generating rate of clusters are denoted as $\lambda_D$ and $\lambda_B$ respectively. The initial number of clusters is calculated as $Nc0=\lambda_{B}/\lambda_D$. In Fig. 4, $\text{IRS(X,Y)}$ means the element on IRS whose index is $\text{(X,Y)}$. The death probability on $x$ direction is calculated as \cite{Bianji}
	\begin{equation} P_{xdeath}=\text{exp}(-\lambda_B\frac{\delta^{I_x}\cos{\beta_E^{I_x}}}{D_C^A}).
	\end{equation}
	where $D_C^A$ is the correlation factor dependent on the scenario. The method of clusters evolution from $\text{IRS(X-1,1)}$ to $\text{IRS(X,1)}$ is generating a set of random variables following the uniform distribution and comparing them with $P_{xdeath}$. If the number is smaller than the death probability, the corresponding cluster is invisible for the next antenna element $\text{IRS(X,1)}$. Then we consider the new cluster generation for $\text{IRS(X,1)}$. The number of new clusters is denoted as $N_x$ which follows the Poisson distribution \cite{3DCM}. The mean value of this variable is calculated as $\frac{\lambda_{B}}{\lambda_{D}}(1-P_{xdeath})$. Then the maximum cluster index of the next element increases by this value compared to the last element. For another direction, the different thing is the number of the uniform distributed variables. The number of the random variables is $M_x \times M_y$. At last, we will obtain a matrix whose dimension is $M_x \times M_y \times Nc0$, containing the information of all the clusters' visbility for every antenna element. 
\section{Statistical Properties and Results analysis}
	In this section, some typical statistical properties of the proposed non-stationary theoretical IRS channel model are derived. 
\subsection{Time Autocorrelation Function}
	The simulated time varying time autocorrelation function between $A_q^B$ and $A_p^U$ can be shown as
\begin{equation}
		R_{qp,n,r,sim}(t,f;\Delta t)=E\left\lbrace h_{qp,n}(t,f)h_{qp,n}^{*}(t+\Delta t,f)\right\rbrace 
\end{equation}
	where $E[\cdot]$ is the expection operator and $(\cdot)^{*}$ denotes the complex conjugate operator. Here, we consider the situation of one cluster whose index is $n$ and one reflecting element whose index is $r$. Also, The analytical result can be written as
\begin{equation}
\begin{split}
	&R_{qp,n,r,ana}(t,f;\Delta t)\\
	=&R_{qr,n,ana}(t,f;\Delta t)\times R_{rp,n,ana}(t,f;\Delta t)\\
	&\times e^{j\theta_{r}(t)-j\theta_{r}(t+\Delta t)}
\end{split}	
\end{equation}	
	where the analytical result of the subchannel from $A_q^B$ and $A_r^I$ is calculated as
\begin{equation}
\begin{split}
	&R_{qr,n,ana}(t,f;\Delta t)\\
	=&\sum_{m_{n}=1}^{M}[P_{qr,m_n}(t)P_{qr,m_n}(t+\Delta t)]^{1/2}\\
	&e^{j\frac{2\pi(f_c-f)}{c}[d_{qr,m_n}(t)-d_{qr,m_n}(t+\Delta t)]}.
\end{split}
\end{equation}
	When there exists a LoS component and we only consider the $n\text{th}$ path of NLoS component, it is calculated as
\begin{equation}
\begin{split}
	R_{qp,r}(t,f;\Delta t)=&\frac{K}{K+1}R_{qp,r}^{L}(t,f;\Delta t)\\
	&+\frac{1}{K+1}R_{qp,n,r}^{N}
\end{split}
\end{equation}
	where $R_{qp,r}^{L}(t,f;\Delta t)$ is the autocorrelation function of the LoS component and $K$ is the Rician factor. And the LoS component of the autocorrelation function is calculated as
\begin{equation}
\begin{split}
	&R_{qp,r}^{L}(t,f;\Delta t)\\
	=&[P_{qp}^{L}(t)P_{qp}^{L}(t+\Delta t)]^{1/2}e^{j\frac{2\pi(f_c-f)}{c}[d_{qp.r}^{L}(t)-d_{qp,r}^{L}(t+\Delta t)]}\\
	&e^{j\theta_{r}(t)-j\theta_{r}(t+\Delta t)}
\end{split}
\end{equation}	

	Then we can get the simulation result and the analytical result shown in Fig. 5. The fact that the two curves do not coincide shows the time-nonstationarity.
\begin{figure}[tb]
	\centerline{\includegraphics[width=0.5\textwidth]{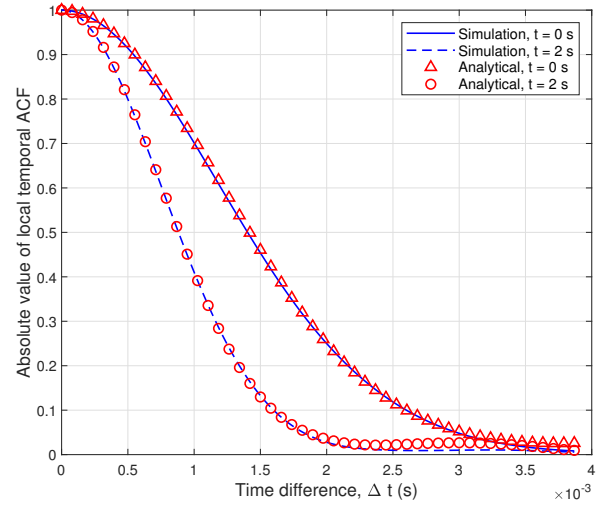}}
	\caption{The comparison of time autocorrelation function between the simulation and the analytical at $t$=0 s and $t$=2 s. ($D_{BI}$=100 m, $D_{IU}$=200 m, $f_c$=62 GHz, $v^B$=10 m/s, $v^U$= 10 m/s, $q$=1, $p$=1, $r$=1, $n$=1)}
	\label{fig_7}
\end{figure}
	Fig. 6 shows that the value of the autocorrelation function is higher after we use IRS. It means that we can control the channel better with the aid of IRS.
\begin{figure}[tb]
	\centerline{\includegraphics[width=0.5\textwidth]{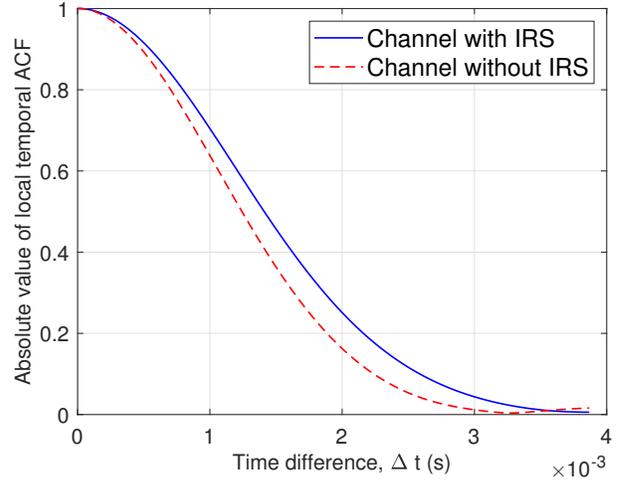}}
	\caption{The comparison of time autocorrelation function between the channel with IRS and without IRS. ($v^B$=10 m/s, $v^U$= 10 m/s, $D_{BI}$=100 m, $f_c$=62 GHz)}
	\label{fig_8}
\end{figure}
\subsection{Spatial Cross-correlation Function}
	The simulation result of the spatial cross-correlation function is 
\begin{equation}
	R_{qp,qp',n,sim}(t,f;\Delta r)=E\left\lbrace H_{qp}(t,f)H_{qp'}^{*}(t,f)\right\rbrace. 
\end{equation}
	The analytical result is calculated as 
\begin{equation}
\begin{split}
	R_{qp,qp',n,ana}(t,f;\Delta r)=&\sum_{m_n=1}^{M}[P_{qp,m_n}(t)P_{qp',m_n}(t)]^{\frac{1}{2}}\\
	&e^{j\frac{2\pi(f_c-f)}{\lambda f_c}[d_{qp,m_n}(t)-d_{qp',m_n}(t)]}.
\end{split}
\end{equation}
	
	As shown in Fig. 7, we can see that two curves are almost the same because the antenna interval is small due to the high frequency.
\begin{figure}[tb]
	\centerline{\includegraphics[width=0.5\textwidth]{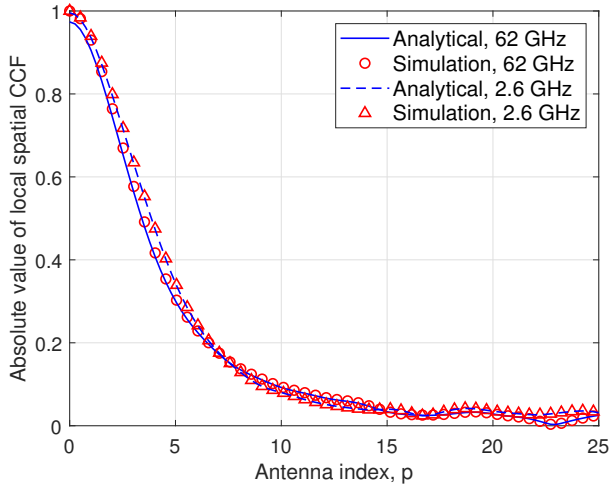}}
	\caption{The comparison of spatial cross-correlation function between the simulation and the analytical at 62 GHz and 2.6 GHz.($M_B$=100, $M_U$=1, $D_{BU}$=100 m, $v^B$=10 m/s, $v^U$= 10 m/s)}
	\label{fig_9}
\end{figure}
\subsection{RMS Delay Spead CDF}
	To get the CDF of RMS delay spread, we should operate the channel simulation many times and count these results then we can get the RMS delay spread CDF. The RMS delay spread of the channel from BS to IRS can be calculated as
\begin{equation}
	DS_{BI}=\sqrt{\sum_{m_n=1}^{M_n}P_{m_n}(\frac{d_{m_n}^B+d_{m_n}^I}{c})^2-(\sum_{m_n=1}^{M_n}P_{m_n}\frac{d_{m_n}^B+d_{m_n}^I}{c})^2}.
\end{equation}
	Then, we should calculate it for another channel and add them two together. As shown in Fig. 8, we can see that under the condition of more dispersed scatters, the delay spread is bigger.
\begin{figure}[tb]
	\centerline{\includegraphics[width=0.5\textwidth]{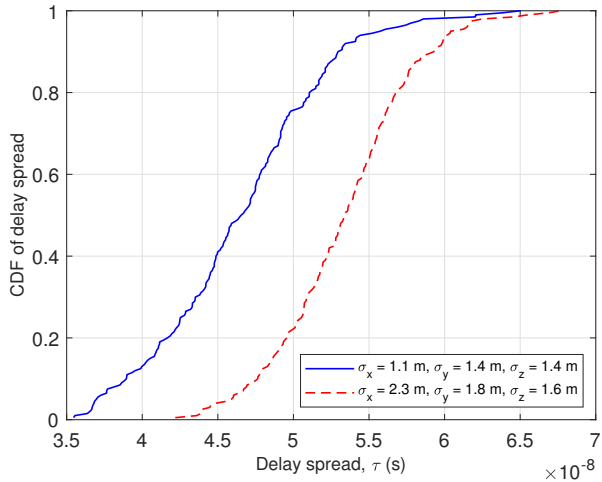}}
	\caption{The comparison of RMS delay spread CDF between the different variance of the scatterers' coordinate values.}
	\label{fig_10}
\end{figure}
\section{Conclusions}
In this paper, a GBSM for IRS-based 6G channel has been proposed. The statistical properties such as time ACF, spatial CCF, local doppler spread, and DS CDF have been simulated and analysed. The good agreement between simulation results and analytical results illustrates the correctness of the proposed channel model. Temporal non-stationary properties of the proposed channel model can be seen from the difference among the curves. The non-stationarities of time and space are caused by time and space varying distances and cluster evolution matrices. The fact that time ACF's value of the situation using IRS is higher than it without IRS illustrates that IRS can seperate the channel and change the statistical properties of the channel.

\section*{Acknowledgment}
\small {This work was supported by the National Key R\&D Program of China under Grant 2018YFB1801101, the National Natural Science Foundation of China (NSFC) under Grant 61960206006, the Frontiers Science Center for Mobile Information Communication and Security, the High Level Innovation and Entrepreneurial Research Team Program in Jiangsu, the High Level Innovation and Entrepreneurial Talent Introduction Program in Jiangsu, the Research Fund of National Mobile Communications Research Laboratory, Southeast University, under Grant 2020B01, the Fundamental Research Funds for the Central Universities under Grant 2242020R30001, the Huawei Cooperation Project, and the EU H2020 RISE TESTBED2 project under Grant 872172.}

\end{document}